\documentclass[12pt, modern]{aastex631}
\usepackage{graphicx}
\usepackage{pgfplots}
\usepackage{longtable}

\shorttitle{Pisgah Astronomical Research Institute}
\shortauthors{Shelby, Scharlach, Matejic, Everett, Morgan}

\graphicspath{{./}{figures/}}

\pgfplotsset{compat=1.17}

\begin{document}

\title{An Optical Analysis of Sunspots as Predictors of Geomagnetic Storms}

\author{Matthew Shelby, Scott Scharlach, Petar Matejic, RJ Everett, Colton Morgan}
\affiliation{Pisgah Astronomical Research Institute \\
1 PARI Drive \\
Rosman, NC 28772, USA}

\begin{abstract}

Although a variety of phenomena may create a geomagnetic storm on Earth, the most severe geomagnetic storms arise from solar activity, and in particular, coronal mass ejections (CMEs) and solar flares. CMEs and flares originate primarily from sunspots. The ‘‘aa index'' is a metric which ranks all of the strongest geomagnetic storms between 1868 and 2010 based on a variety of characteristics taken from several sources. This paper examines correlations between the aa index of the most severe geomagnetic storms and the intrinsic characteristics of the sunspots from which they originated. We find a correlation between the total rank of the aa index of the storms and the ‘‘total intensity'' of the sunspot, where total intensity is defined as the sunspot's mean intensity multiplied by its area. The correlation has an R$^2$ = 0.690 and R$^2$ = 0.855 when a potentially corrupted data point is removed.

\end{abstract}

\keywords{Sunspots -- Geomagnetic Storms -- Coronal Mass Ejections -- aa Index}

\section{Introduction} \label{sec:intro}

When the Earth’s magnetosphere is sufficiently disturbed, a geomagnetic storm can potentially arise (Space Weather Prediction Center [Accessed September 23, 2021]). These disturbances mostly emerge from solar flares and Coronal Mass Ejections (CMEs) which are oriented towards Earth. These CMEs are composed of magnetized plasma and can form in minutes. When such structures occur, they are violently discharged from the Sun and sent out into the solar system (Temmer, M. 2021).

Such an event may give rise to a geomagnetic storm on Earth. The effects of a geomagnetic storm can be devastating, such as causing entire nations' power grids to fail. However, geomagnetic storms seem to be very rare. Studies have shown that space weather events that originate from the sun occur mostly during the solar maximum phase (Hayakawa, H. et al., 2020); therefore, space weather is not as prevalent during the solar minimum phase.

Strong geomagnetic storms may still occur during the solar minimum, such as the geomagnetic storm of 1903 (Hayakawa, H. et al., 2020). A detailed study was conducted by (Loewe, C.A. and G.W. Prölss, 1997) which found that the number of great storms (those with $<$ -350 nT) occur once every 6 years while 3 strong ($<$ -100nT) and 2 severe ($<$ -200nT) storms occur every year. Due to the large number of geomagnetic storms which occur annually, as well as the occurrence of massive storms such as the Carrington event of 1859, it is crucial to study these storms and their effects on Earth so that solar physicists may more accurately predict when the next geomagnetic storm will occur. 

There are many factors to consider when predicting whether a sunspot will flare up, such as its background magnetic field strength and spectra (Su, Y. et al., 2007), (Lites, B.W. et al., 1993). Large and particularly complex structured sunspots have been known to flare up in the past as well (Hayakawa, H. et al., 2021). With the Solar Dynamics Observatory (SDO), we now have a telescope in space that can see the sun in several key wavelengths to measure and predict the sun's activity. Yet new technology and theories about predicting solar activity are still imperfect. Merely one geomagnetic storm as severe as the Carrington event of 1859 is capable of causing the entire world's power grid to fail, which may result in trillions of dollars worth of damage. The aim of this paper is to discover whether or not the ranks of the geomagnetic storms of the aa index are correlated with values analyzed from the sunspots from which these storms originated. If such a correlation exists, this may give insight into methods that can predict if a sunspot, or group of sunspots, will flare up and pose a threat to Earth. It is important to analyze past events to acquire a better grasp on when the next major geomagnetic storm will arise.

\section{Procedure} \label{sec: Procedure}

In order to analyze sunspots prior to severe geomagnetic storms, images of the sun taken 0-4 days before the occurrence of geomagnetic storms were required. The images of the sun were acquired from two databases: the Astronomical Photographic Data Archive (APDA) and the Kodaikanal Solar Observatory (KSO). APDA is the second largest collection of astronomical plates in the world and is housed at the Pisgah Astronomical Research Institute (PARI) in North Carolina, United States. The images from both APDA and KSO consisted of digital scans of emulsion plates; emulsion plates are thin glass squares which have a negative image imprinted on it made from dried chemicals. 

Approximately 0-4 days pass between coronal mass ejections leaving the sun and arriving at earth (Temmer, M., 2021). Therefore, we analyzed the plates of the sun corresponding to dates within that time window of the focus event. The APDA emulsion plates were scanned using a Epson Expression 11000XL Photo Scanner while the KSO plates were downloaded from KSO’s online database as .TIFF files. Our project is concerned only with possible correlations between Coronal Mass Ejections (CMEs) and solar flares with geomagnetic storms; therefore, geomagnetic storms which were caused by phenomena other than CMEs and solar flares were not studied. 

Some emulsion plates taken prior to geomagnetic storms were in poor condition or not available in APDA or KSO. Therefore, those geomagnetic storms were not studied despite their status as some of the strongest geomagnetic storms ever recorded. Some of the plates that were scanned and analyzed were 118 years old at the time of digitization, and it cannot be assured that the plates are in optimal condition. Once a list of relevant emulsion plates with associated focus dates was compiled, we searched APDA and KSO for all the dates that were 0-4 days before known geomagnetic storms. In an effort to reduce randomized error, only geomagnetic storms which corresponded to at least three good-condition emulsion plates were studied; events in which only one or two good-condition plates were available were not studied.

The aa index rank for a geomagnetic storm accounts for a variety of factors including the peak aa value, storm duration, and number of peaks in the storm interval (Vennerstrom et al. 2016). Because of the high number of variables, definitively ranking the severity of a storm poses a challenge. The aa index rankings are intended to serve primarily as a general guideline but not a comprehensive analysis (Vennerstrom et al. 2016). The aa index contains data beginning in 1868, while the Disturbance Storm Time (dst) records contain data beginning in 1957. For our study, the aa index was chosen because our earliest emulsion plates are from 1903, before the earliest dates in the dst (Hayakawa, H. et al., 2021). In contrast, prior studies provide dst \emph{estimates} of pre-1957 strong geomagnetic storms, such as the 1903 storm (Hayakawa, H. et al., 2020), but there are no estimates for every storm before 1957. 

The digitized KSO and APDA emulsion plates were all analyzed using the program AstroImageJ. For each plate, the following values were recorded: the angle between the sunspot and Earth relative to the center of the sun (\emph{$\theta$}), the apparent mean intensity and maximum intensity of the sunspot (Mean \emph{$I_{\theta}$} and Max \emph{$I_{\theta}$}, respectively), the mean intensity and maximum intensity of the sunspot with limb darkening taken into account (Mean \emph{$I_{0}$} and Max \emph{$I_{0}$}, respectively, or ‘‘Mean Head-On Intensity'' and ‘‘Max Head-On Intensity,'' respectively), the observed area of the sunspot (\emph{$A_{obs}$}), the area of the sunspot with foreshortening taken into account (\emph{$A_{0}$}, or ‘‘head-on Area''), and the observed percentage of the solar surface that the sunspot covered (\emph{\% Surf}).

This study also determined the orientation of the sun for each emulsion plate. For each event, the relative location of the sunspot at the earliest date in the event was compared to the relative location of the sunspot at the latest date in the event. The sun rotates West to East as observed from earth; therefore the proper orientation of the emulsion plates is the orientation in which the sunspot's relative location moves from West to East.

\begin{figure}
    \centering
    \includegraphics[width=0.5\textwidth]{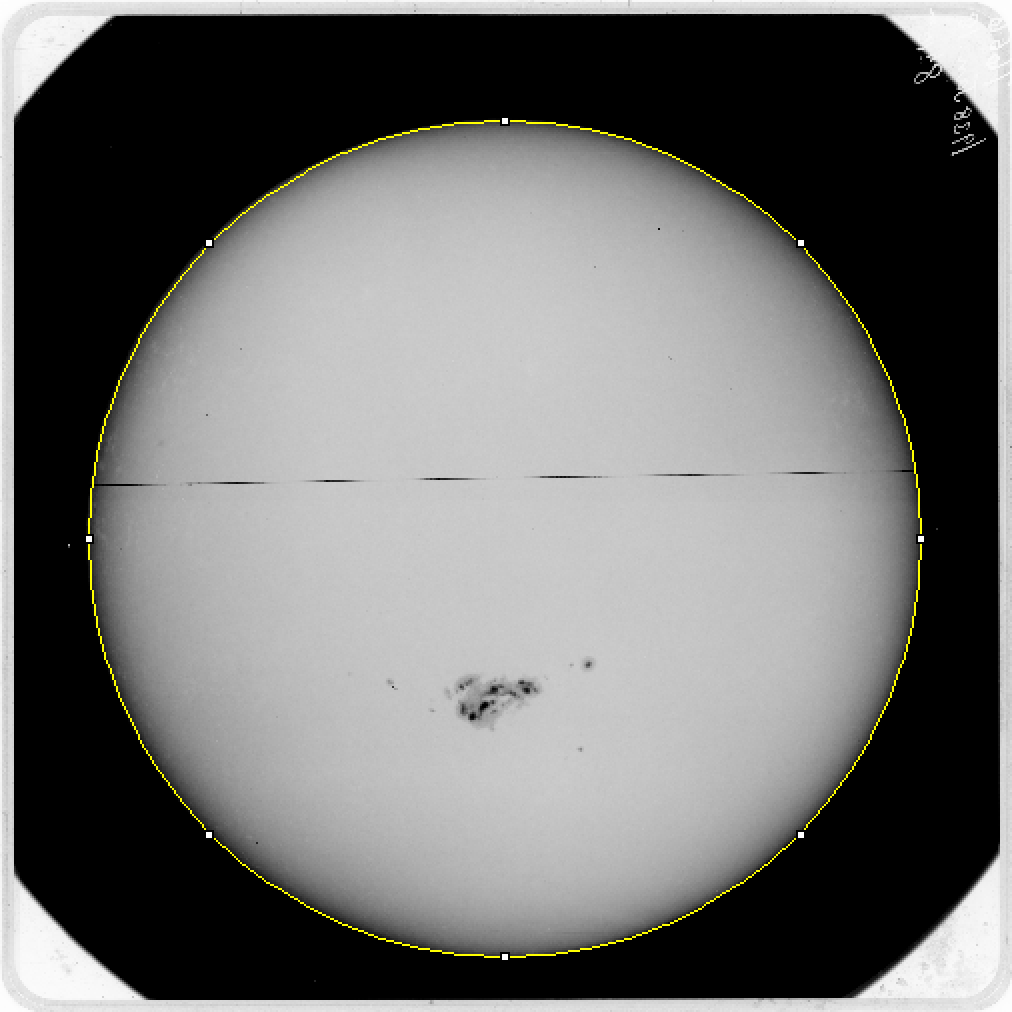}
    \caption{This is a solar plate from January $18^{th}$ 1938. On the Western and Eastern edges of the ellipse are two squares which indicate the \emph{y}-coordinate of the center of the sun. Similarly, the Northern and Southern edges indicate the  \emph{x}-coordinate of the center of the sun. The fitted circle also provided the area and diameter of the sun in pixels. Using the known radius of the sun $R_{Solar}$ = 1.3927*$10^6$ km, we established a rate of conversion between pixels and kilometers. The position of the sunspot’s center ($x_2$, $y_2$) was also found with AstroImageJ.}
    \label{fig: Emulsion Plate}
\end{figure}

A calculation of the angle between the sunspot and Earth relative to the center of the sun, \emph{$\theta$}, requires a knowledge of the pixel location (\emph{x,y}) of the center of the sun on the digitized emulsion plate. To find the center of the sun, the Ellipse tool in AstroImageJ was used to fit a circle precisely (to the nearest pixel) over the circumference of the sun. This indicated two points on the circle which possessed the same x-coordinate as the center of the sun and two points which possessed the same y-coordinate as the center of the sun, allowing us to deduce the position ($x_1$, $y_1$) of the center (Figure \ref{fig: Emulsion Plate}).

The angle of the sunspot was calculated with the equation $\theta$ = $sin^{-1}$(d/$R_{Solar}$), where d is the distance between the center of the sun and the sunspot (with units of pixels) and $R_{Solar}$ is the radius of the sun (in units of pixels). 

The value of \emph{d} was calculated with the distance formula, 
\begin{equation}
   d =  \sqrt{(x_1 - x_2)^2 + (y_1 - y_2)^2}.
\end{equation}

The observed maximum pixel intensity of the sunspot, observed mean pixel intensity of the sunspot, and observed area of the sunspot were recorded using the Maximum Value, Mean Value, and Area functions in AstroImageJ, respectively.

\begin{figure}
    \centering
    \includegraphics[width=0.5\textwidth]{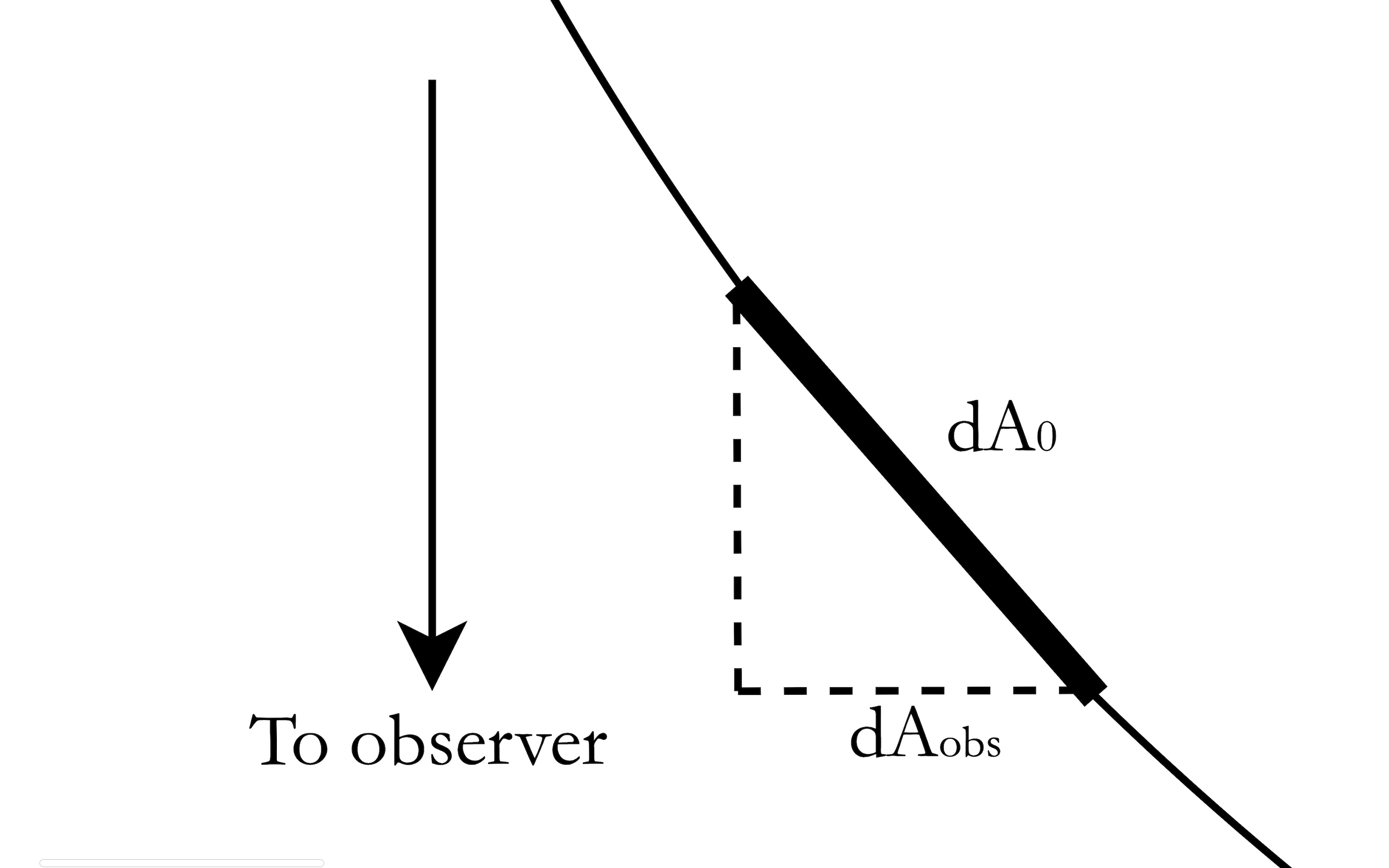}
    \caption{The sunspot is positioned at an angle \emph{$\theta$} relative to the observer. Therefore, the area covered in the digitized emulsion plate is less than its true area, or ‘‘head-on area.'' An infinitesimal area \emph{$dA_{0}$} on the surface of the sun can be approximated as a flat surface. Because the area of the sunspot is small compared to the area of the surface of the sun, we use this approximation (to an accuracy of 10\%). Note that the angle between \emph{$dA_{obs}$} and \emph{$dA_{0}$} is equal to \emph{$\theta$}.}
    \label{fig: Area}
\end{figure}

The area of the sunspot with foreshortening taken into account was found using the equation $dA_{obs}$ = $dA_0$ $\cos(\theta)$, where $dA_{obs}$ is an infinitesimal two-dimensional segment of the observed area, $dA_0$ is an infinitesimal segment of the area with foreshortening accounted for, and $\theta$ is the angle between the sunspot and observer with respect to the center of the sun (Figure \ref{fig: Area}). The sunspots were sufficiently small compared to the surface area of the sun that $dA_{obs}$ $\approx$ $A_{obs}$ and $dA_0$ $\approx$ $A_0$. The observed percentage of the solar surface was calculated by dividing the observed area of the sunspot in pixels by the area of the sun in pixels.
\ref{fig: Area}

\begin{figure}
    \includegraphics[width=0.9\textwidth]{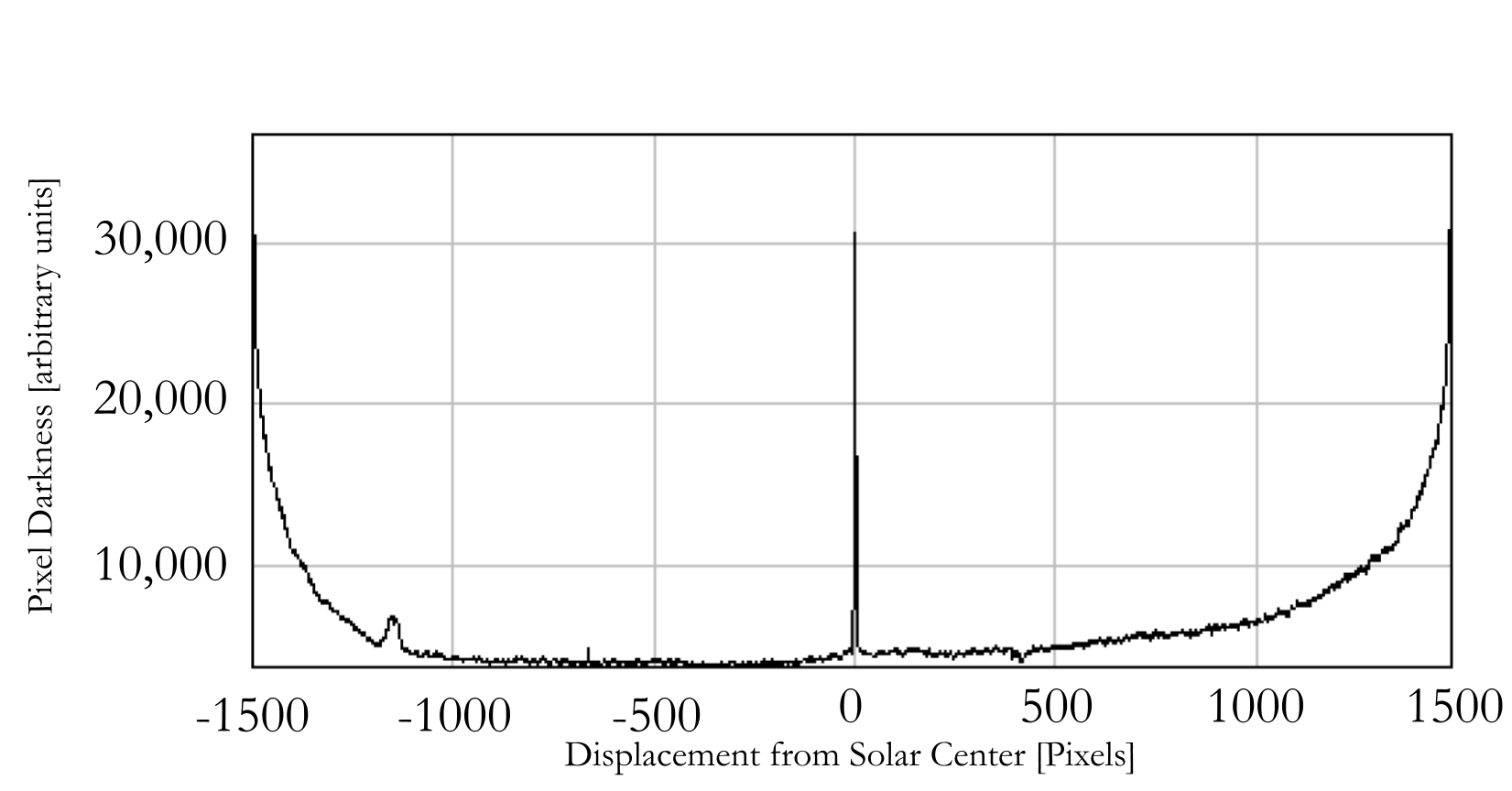}
    \caption{A displacement-to-darkness graph across the diameter of the sun on September 20th, 1909. The local maximum at approximately -1200 pixels represents the sunspot. The observed intensity \emph{$I_{\theta}$} measures the intensity of the sunspot relative to the x-axis (Pixel Darkness = 0). In contrast, the head-on intensity \emph{$I_{0}$} measures the intensity of the sunspot relative to the intensity of the solar surface that would otherwise be at that displacement. The sharp peak at Displacement = 0 is due to a black line drawn onto the emulsion plate to indicate the solar equator; it has no physical significance.}
    \label{fig: Limb_Darkening}
\end{figure}

One of the greatest challenges of this study was properly taking into account limb darkening. Because the sun is a sphere and is of non-uniform density, the center of the sun appears brighter to an observer than the edges, or limbs, of the sun. This effect is known as limb darkening (Figure \ref{fig: Limb_Darkening}). Because the sunspots closer to the limbs of the sun are surrounded by a darker background than the sunspots closer to the center, a lack of taking limb darkening into account would skew our results significantly.

To account for limb darkening, this study used the equation,
\begin{equation}
    \frac{I_{\theta}}{I_0}=a_0+a_1cos{\theta}+2a_2cos{\theta}^2
\end{equation}
This equation is taken from Introduction to stellar astrophysics, Böhm-Vitense 1989 (p. 233), where $I_{\theta}$ is the apparent brightness of a point on the sun, $I_{0}$ is the brightness of the sun at the center, \emph{$a_{0}$}, \emph{$a_{1}$}, and \emph{$2a_{2}$} are empirically observed constants provided in (Böhm-Vitense 1989), and ${\theta}$ is the angle between the sunspot and the observer with respect to the center of the sun. At 501 nm,
\begin{equation}
   ({a_0},{a_1},{2a_2})=(0.2593,0.8724,-0.1136)
\end{equation}
This paper deals only with the relative brightness of pixels compared to each other, and thus our equation is unitless.
Because this study dealt with the darkness of sunspots relative to the brightness of the sun, we modified our equation to be
\begin{equation}
    \frac{I_0}{I_{\theta}}=a_0+a_1cos{\theta}+2a_2cos{\theta}^2
\end{equation}
so that $I_0$ and $I_{\theta}$ measure the pixel intensity of darkness, not brightness.
To avoid errors and discrepancies, every plate was analyzed by two different people to make sure that the method being used was consistent. If two people were getting completely different numbers, a third party was brought in to analyze the plate. 

\section{Results and Discussion} \label{sec: Results_And_Discussion}

Following the collection of all relevant data from the solar plates, we attempted to find a correlation between sunspots and geomagnetic storms by comparing certain values from each. For the geomagnetic storms, the primary focus was on the overall rank of the storm, while a secondary focus was placed on peak aa value. While the overall rank of the storm does not directly correlate with any single observable value, we found it to be the best indication of the strength of the storm as a whole. No singular value could encompass all of the data. We also examined the peak aa  value to determine whether a correlation existed between the peak intensity of the sunspot and the peak aa value. 

For the sunspots, this paper focused primarily on the ‘‘total intensity,'' here defined as the area of the sunspot multiplied by its mean intensity. We examined both the observed intensity and head-on intensity, as defined in Section \ref{sec: Procedure}. We also examined the peak intensity of the sunspot, as it appeared to be plausible that the peak intensity of the sunspot might correlate with the peak aa index or perhaps the overall rank. For the values relating to the sunspots, we chose the peak value for the entire event. If there were three dates that were analyzed, only the day with the highest peak intensity or overall intensity was used as this was most likely when the flare or CME occured (See the Appendix, Section \ref{Appendix}, for daily values). 

When looking at the measured peak intensity of the sunspots, we graphed the values for each event against both the peak aa index and the overall rank of the geomagnetic storm. There was not a correlation between peak intensity and either value. The peak intensity and overall rank did not correlate strongly enough to warrant any confident conclusions about the relation between sunspots and geomagnetic storms. The comparison of peak intensity to peak aa index likewise had no statistically significant correlation. We note that there existed several storms had approximately equal aa indices, but few geomagnetic storms had approximately equal intensities. Indeed, the peak aa index of geomagnetic storms correlated with virtually no other characteristic which we examined.

\begin{figure}
\begin{center}
\begin{tikzpicture}[scale=0.75]
\begin{axis}[
    title={Peak Intensity vs. Rank},
    xlabel={Rank},
    ylabel={Peak Intensity},
    xmin=0, xmax=15,
    ymin=0, ymax=60000,
    legend pos=north east,
    ymajorgrids=true,
    grid style=dashed,
]

\addplot+[
    only marks,
    color=black,
    mark=o,
    ]
    coordinates {
    (6,43906)(3,51837)(1,25800)(4,33295)(5,50006)(7,11032)(13,12757)(2,17142)
    };
    
\end{axis}
\end{tikzpicture}
\qquad
\begin{tikzpicture}[scale=0.75]
\begin{axis}[
    title={Peak Intensity vs. aa Peak},
    xlabel={aa Peak},
    ylabel={Peak Intensity},
    xmin=500, xmax=750,
    ymin=0, ymax=60000,
    legend pos=north east,
    ymajorgrids=true,
    grid style=dashed,
]

\addplot+[
    only marks,
    color=black,
    mark=o,
    ]
    coordinates {
    (658,43906)(658,51837)(680,25800)(656,33295)(526,50006)(656,11032)(568,12757)(715,17142)
    };
    
\end{axis}
\end{tikzpicture}
\end{center}
\caption{The two graphs above depict the relationship between the peak intensity of sunspots and both the rank and aa peak of the resulting geomagnetic storms. There is no statistically significant correlation present in either graph.}
\end{figure}
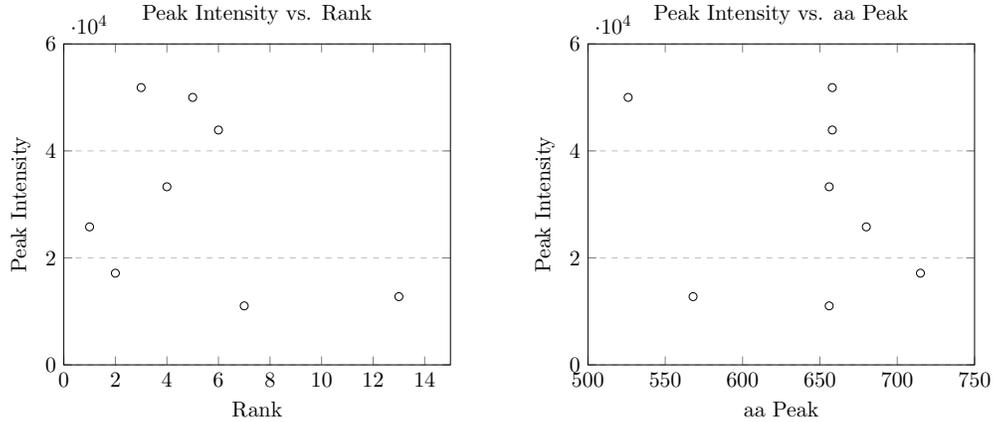

The other values of sunspots which we compared to geomagnetic storms was the total measured intensity of the sunspot. We modelled the overall intensity of a sunspot by multiplying the mean intensity of the sunspot by its total area. Both the overall intensity of the sunspot and the overall rank of the geomagnetic storm provide an overall (but imperfect) indicator of their significance, and thus we found it plausible that the two values might correlate. We utilized three slightly different definitions of total intensity, all of which we compared with the geomagnetic storm's overall rank: the ‘‘observed total intensity'' ($A_{obs}$ * Mean $I_{obs}$), the ‘‘head-on total intensity'' ($A_0$ * Mean $I_0$), and a hybrid of the previous two values ($A_{obs}$ * $I_0$). The highest total intensities for each sunspot were graphed in relation to the overall ranks of the resulting geomagnetic storms. 

For each graph, both a linear and exponential fit was tested. An exponential curve was a better fit for each of the graphs, a conclusion that is not surprising for two main reasons. Firstly, the gaps in aa indices between the storms' ranks are unlikely to be uniform; e.g. the gap between medium-sized geomagnetic storms should be smaller than the gap between large ones. Secondly, a linear regression would eventually lead to negative intensity values, which are not possible. As a result, an exponential fit was used to test the correlation between total intensities and rank.

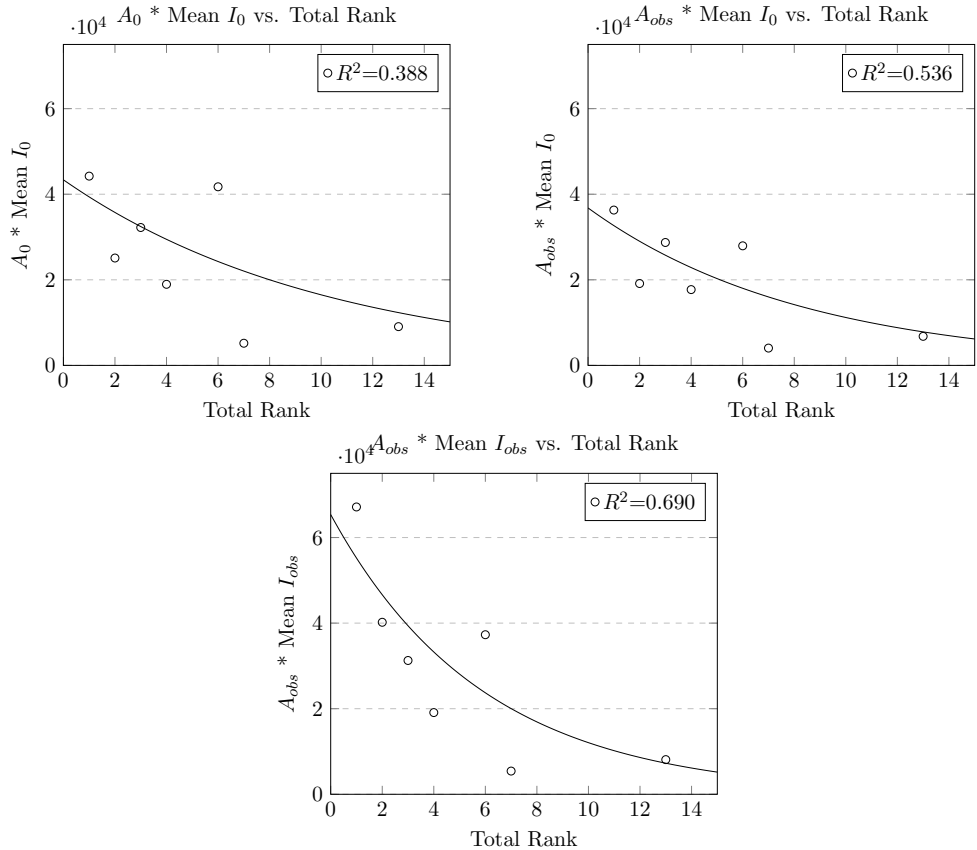
\begin{figure}
\begin{center}
\begin{tikzpicture}[scale=0.75]
\begin{axis}[
    title={$A_0$ * Mean $I_0$ vs. Total Rank},
    xlabel={Total Rank},
    ylabel={$A_0$ * Mean $I_0$},
    xmin=0, xmax=15,
    ymin=0, ymax=75000,
    legend pos=north east,
    ymajorgrids=true,
    grid style=dashed,
]

\addplot+[
    only marks,
    color=black,
    mark=o,
    ]
    coordinates {
    (1,44230)(4,18948)(7,5178)(13,9056)(2,25092)(6,41737)(3,32223)
    };
    
\addplot [
    domain=0:15, 
    samples=100, 
    color=black,
    ]
    {43361*e^(-0.0967*x)};
\addlegendentry{$R^2$=0.388}
    
\end{axis}
\end{tikzpicture}
\qquad
\begin{tikzpicture}[scale=0.75]
\begin{axis}[
    title={$A_{obs}$ * Mean $I_0$ vs. Total Rank},
    xlabel={Total Rank},
    ylabel={$A_{obs}$ * Mean $I_0$},
    xmin=0, xmax=15,
    ymin=0, ymax=75000,
    legend pos=north east,
    ymajorgrids=true,
    grid style=dashed,
]

\addplot+[
    only marks,
    color=black,
    mark=o,
    ]
    coordinates {
    (1,36299)(4,17711)(7,4044)(13,6792)(2,19133)(6,27942)(3,28721)
    };
    
\addplot [
    domain=0:15, 
    samples=100, 
    color=black,
    ]
    {36793*e^(-0.119*x)};
\addlegendentry{$R^2$=0.536}
    
\end{axis}
\end{tikzpicture}

\begin{tikzpicture}[scale=0.75]
\begin{axis}[
    title={$A_{obs}$ * Mean $I_{obs}$ vs. Total Rank},
    xlabel={Total Rank},
    ylabel={$A_{obs}$ * Mean $I_{obs}$},
    xmin=0, xmax=15,
    ymin=0, ymax=75000,
    legend pos=north east,
    ymajorgrids=true,
    grid style=dashed,
]

\addplot+[
    only marks,
    color=black,
    mark=o,
    ]
    coordinates {
    (1,67144)(4,19096)(7,5425)(13,8100)(2,40196)(6,37288)(3,31267)
    };
    
\addplot [
    domain=0:15, 
    samples=100, 
    color=black,
    ]
    {65366*e^(-0.169*x)};
\addlegendentry{$R^2$=0.690}
    
\end{axis}
\end{tikzpicture}
\end{center}
\caption{The three graphs above depict the relationship between the overall rank of the geomagnetic storms and several different methods for evaluating the overall intensity of the preliminary sunspot.  An exponential regression, marked by the black line, is present in each of the graphs. The resulting correlation between rank and each of the three measures of overall intensity is depicted by an $R^2$ value.}
\end{figure}

As shown by the graphs, the total intensity measure that had the highest correlation to the overall rank of the storm was the observed total intensity. The $R^2$ value of the exponential fit was higher than the other two at 0.690. When looking at the graph of observed total intensity vs. rank, it is clear that the rank six storm, which occurred in 1903, is somewhat of an outlier. Because 1903 was the oldest date that we analyzed, it is possible that those plates were in the worst condition and their data was somewhat corrupted. It is also possible that the 1903 storm is an outlier for some reason we have not found. 

\begin{figure}
\begin{center}
\begin{tikzpicture}[scale=0.75]
\begin{axis}[
    title={$A_{obs}$ * Mean $I_{obs}$ vs. Total Rank},
    xlabel={Total Rank},
    ylabel={$A_{obs}$ * Mean $I_{obs}$},
    xmin=0, xmax=15,
    ymin=0, ymax=75000,
    legend pos=north east,
    ymajorgrids=true,
    grid style=dashed,
]

\addplot+[
    only marks,
    color=black,
    mark=o,
    ]
    coordinates {
    (1,67144)(4,19096)(7,5425)(13,8100)(2,40196)(3,31267)
    };
    
\addplot [
    domain=0:15, 
    samples=100, 
    color=black,
    ]
    {67072*e^(-0.215*x)};
\addlegendentry{$R^2$=0.855}
    
\end{axis}
\end{tikzpicture}
\end{center}
\caption{The graph above depicts the relationship between rank and the total intensity value that had the best correlation, observed total intensity. In this graph, however, the sixth ranked storm, an outlier, was removed to show the correlation without it. It is possible that this graph shows a more accurate relationship between rank and observed total intensity.}
\end{figure}
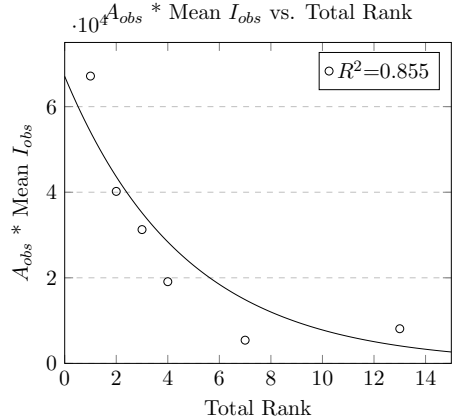

When that data point is removed from the graph, the correlation becomes even stronger. With an $R^2$ value of 0.855, this exponential fit has a strong correlation, suggesting a legitimate connection between rank and observed total intensity. This correlation constitutes the main result of this work.

We find it conceptually unsurprising that the aa index rank correlated more closely to the observed total intensity than the head-on total intensity or hybrid model. Although the head-on values of intensity and area provide useful information about the true nature of the sunspot, the geomagnetic storms which resulted from the sunspots occurred here on Earth. Given that the Earth was not always directly above the sunspot we analyzed, the strength of geomagnetic storms would not result from the head-on intensity of the sunspots. Only a fraction of the emissions would be pointed in the direction of Earth, and thus only a fraction of the total intensity would be involved in causing a geomagnetic storm. Therefore, using the observed characteristics of the sunspot to find a correlation might provide a better representation of how it affects Earth, while the head-on values of the sunspot are important for characterizing it.

\section{Conclusion} \label{sec: Conclusion}

In this paper, we aimed to find a correlation between the rank of a geomagnetic storm from the aa index to values analyzed from the sunspots they came from. A correlation was found between the rank and the total intensity of the sunspots ($A_{obs}$ * Mean $I_{obs}$) with an $R^2$ value of .690 when including the outlier from 1903 and an $R^2$ 0.855 when excluding it. Further insight on these storms may be gained from analyzing the physical structure of the sunspots, spectroscopic analysis, and measuring its background magnetic field strength. Analyzing more data of geomagnetic storms that were caused by solar flares and CMEs would prove useful as there are some dates that were missing. This would entail analyzing events that rank lower on the overall aa index to see if it fits with the correlation we have found. 

We conclude that the aa index rankings are a reasonable measurement of the severity of the most significant geomagnetic storms, as the rankings correlate well with our data and findings. One strong geomagnetic storm has the potential to wipe out an entire country's power grid, interrupt satellite communication, and harm astronauts that are in orbit around the Earth. With the world so heavily reliant on electronics, it may take decades to fully recover from an event like the Carrington event, particularly if there is no proper preparation beforehand. Continuing to study and analyze the connection between solar activity and geomagnetic storms will be highly beneficial to us on Earth. If humans are sent to other bodies in the solar system, such as Mars, predicting geomagnetic storms will be especially important, as the Martian atmosphere is thinner than on Earth and therefore space weather may cause significant harm to the astronauts and electrical components.

A possible future study may involve converting our unitless data in the Appendix (Section \ref{Appendix}) to metric units or cgs units. A knowledge of the exact flux of the sunspots relative to the solar surface may yield insightful results. Future studies may also analyze images of the sun corresponding to dates that this study could not analyze due to broken or missing plates in APDA or KSO.

\section{Acknowledgements}

We thank Thurburn Barker for being the advisor for this project and helping with everything involving APDA. We also would like to thank Sarah Gibson from the High Altitude Observatory at NCAR who helped answer many important questions and provided excellent insights. Thank you to the Pisgah Astronomical Research Institute (PARI) for giving us all the resources necessary to complete this project. Thank you to the Kodaikanal Solar Observatory (KSO) for having a free open database that helped supply the data that was not available in APDA. 

\vfill

\pagebreak

\section{Appendix} \label{Appendix}

\begin{figure}[h]
    \centering
    \includegraphics[width=\textwidth]{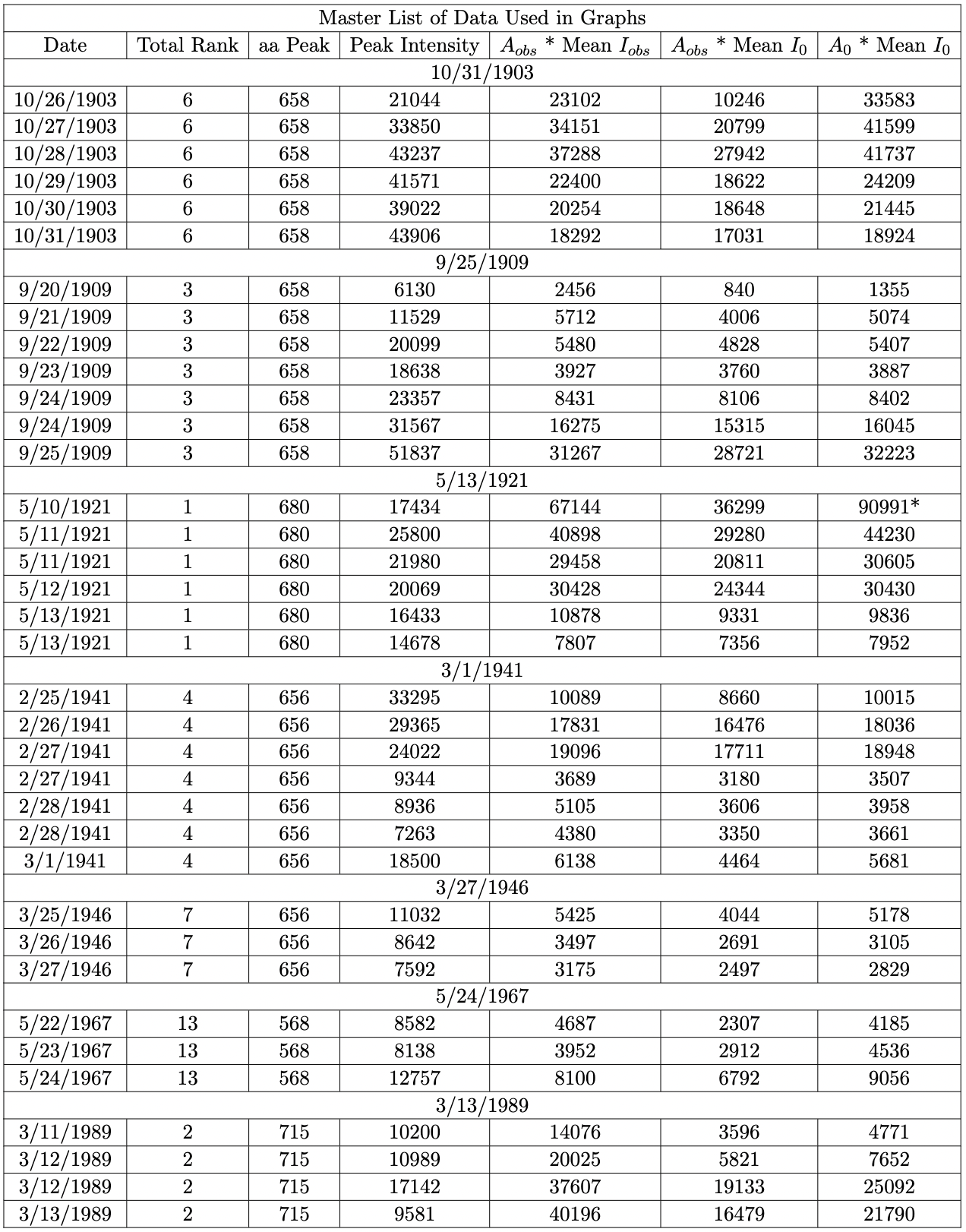}
    \caption{The data used in the figures in Section \ref{sec: Results_And_Discussion} is tabulated here. The data was acquired from digital scans of emulsion plates of the sun from the APDA and KSO databases. The asterisk in the data for 5/10/21 indicates that the sunspot had split into two, and thus presented a difficulty in accurately representing its area. }
\end{figure}

\pagebreak

\section{References}

Alexander, D., Richardson, I.G., $\&$ Zurbuchen, T.H. 2006, A Brief History of CME Science. In: Coronal Mass Ejections. Space Sciences Series of ISSI, vol 21. Springer, New York, NY. doi: \url{https://doi.org/10.1007/978-0-387-45088-9_1} \\

Böhm-Vitense, E. (1989) Introduction to Stellar Astrophysics. Cambridge: Cambridge University Press. doi: \url{https://doi.org/10.1017/CBO9780511623011}
 
Boyd, R.W. (1983). Radiometry and the detection of optical radiation. New York: John Wiley And Sons.\\

Hayakawa, H. et al., 2020. The extreme space weather event in 1903 October/November: An outburst from the Quiet Sun. The Astrophysical Journal, 897(1). doi: 10.3847/2041-8213/ab6a18\\

Hayakawa, H. et al., 2021. The intensity and evolution of the Extreme solar and geomagnetic storms in 1938 January. The Astrophysical Journal, 909(2), p.197. doi: 10.3847/1538-4357/abc427
\\

Lites, B.W. et al., 1993. Stokes profile analysis and Vector magnetic Fields. VI. fine scale structure of a sunspot. The Astrophysical Journal, 418, p.9. doi: 10.1086/173450 \\

Loewe, C.A. and G.W. Prölss, 1997. Classification and mean behavior of magnetic storms. Journal of Geophysical Research: Space Physics, 102(A7), pp.14209–14213. doi: \url{ https://doi.org/10.1029/96JA04020} \\

Space Weather Prediction Center. Accessed 23 September 2021. Geomagnetic storms. National Oceanic and Atmospheric Administration. Available at: \url{https://www.swpc.noaa.gov/phenomena/geomagnetic-storms.} \\

Su, Y. et al., 2007. What determines the intensity of solar flare/cme events? The Astrophysical Journal, 665(2), pp.1448–1459. doi: 10.1086/519679 \\

Temmer, M. Space weather: The Solar Perspective. Living Rev Sol Phys 18, 4 (2021). doi: \url{https://doi.org/10.1007/s41116-021-00030-3} \\

Vennerstrom, S. et al., 2016. Extreme geomagnetic storms – 1868 – 2010. Solar Physics, 291(5), pp.1447–1481. doi: 10.1007/s11207-016-0897-y \\ 

\end{document}